\documentclass[11pt]{amsart}
\usepackage{amscd,amsthm,amssymb,amsfonts,amsmath}
\newfont{\frakturfont}{eufm10 scaled\magstep1}
\newcommand{\frakg}{\mbox{\frakturfont g}}

\newcommand{\frakh}{\mbox{\frakturfont h}}

\newcommand{\p}{\partial}
\newtheorem{theorem}{Theorem}[section]

\newtheorem{define}[theorem]{Provisional Definition}

\title{B fields from a Luddite Perspective}
\author{Mark Stern}
\date{}
\begin{document}

\begin{abstract} In this talk, we discuss the geometric realization of $B$ fields and higher p-form potentials on a manifold $M$ as connections on affine bundles over $M$. We realize D branes on $M$ as special submanifolds of these affine bundles.  As an  application of this geometric understanding of the $B$ field, we give a simple geometric explanation for the Chern-Simons modification of the field strength of the heterotic B field.
 \end{abstract}
 \maketitle
\section{Introduction}
The mathematical description of $B-$fields in terms of gerbes (see for example \cite{H1}) is too abstract to be useful for many basic computations. In this talk we discuss some of our recent work \cite{S2},\cite{S1} developing a simple, geometric representation of B fields and higher $p-$form potentials on a manifold $M$ as connections on various affine bundles over $M$.

Our geometric representation has three ingredients:
\begin{enumerate}
\item the representation of the B field on a manifold $M$ as a connection on an affine bundle $E$ over $M$,
\item a dictionary between string sigma model fields and differential operators on $E$, and
\item the representation of $D$ branes as submanifolds of $E$ (not $M$).
\end{enumerate}

The choice of $E$ depends on the sector of string theory under consideration. Choosing $E$ to be an affine bundle is probably useful only in a low energy approximation. In fact, heterotic string theory already requires more complicated (principal affine) bundles in order to see the full Green-Schwarz mechanism.

Our geometric model of B fields and D branes closely agrees with our stringy expectations. Some easy consequences of this model include
\begin{itemize}
\item A derivation of noncommutative Yang-Mills associated to a B field which suggests generalizations to deformations associated to higher p-form potentials.
\item A geometric representation of topological B type D branes corresponding to coherent sheaves that need not be locally free and that may be "twisted ".
\item A simple geometric explanation for the introduction of the Yang-Mills gauge transformations of B fields in heterotic string theory.
\end{itemize}

We will not discuss noncommutative deformations here, but give a simple example in section 3 of the geometric realization of a topological B type D brane corresponding to a coherent sheaf which is not locally free. In section 4, we show how the geometry of affine bundles leads to the introduction of Yang-Mills gauge transformations of B fields in heterotic string theory.

\section{The Model}
In \cite{S1}, T duality considerations suggested that $B$ fields should be realized as connections on a space of connections. Here we consider a finite dimensional analog: B fields as connections on affine bundles locally modelled on $T^*M$. (See \cite{H2} for related ideas.) First we review some basic facts about connections on $T^*M$.

Local coordinates $(x^i)$ on $M$ define vector fields $\frac{\p}{\p x^i}$ on $M$. These coordinates also determine local coordinates $(x^i,p_s)$ on $T^*M$ which define vector fields $\frac{\p}{\p x^i}$ and $\frac{\p}{\p p_s}$ on $T^*M$. Here we are abusing notation in a standard way, using the same symbol, $\frac{\p}{\p x^i}$, to denote two different vector fields, one on $M$ and the other on $T^*M$. The coordinates thus define a lifting of vector fields on $M$ to vectorfields on $T^*M$ given in a notationally confusing manner as $\frac{\p}{\p x^i}\rightarrow\frac{\p}{\p x^i}.$ This lift is obviously coordinate dependent. The coordinate dependence may be removed by introducing a connection. For example, with the Levi Civita connection, we have the globally well defined lift given by

\begin{equation}\label{eq1}
\frac{\p}{\p x^i}\rightarrow \frac{\p}{\p x^i} + p_n\Gamma^n_{is}\frac{\p}{\p p_s}.
\end{equation}

We now break the vector space structure on $T^*M$ to an affine structure by introducing new allowed local coordinate transformations
$$(x,p)\rightarrow (x,p+\lambda(x)), $$
where $\lambda$ is a locally defined 1 form on $M$. Then (\ref{eq1}) is no longer coordinate independent. To fix this we introduce a 1 form $\mu$ and a 2 form $b$ and define
\begin{equation}\label{eq2}
\frac{\p}{\p x^i}\rightarrow \frac{\p}{\p x^i} + (\mu_{i;s} + b_{is} + p_n\Gamma^n_{is})\frac{\p}{\p p_s}.
\end{equation}
This lift is coordinate independent if
$$\mu\rightarrow \mu+\lambda, \mbox{  and  }  b\rightarrow b+d\lambda, \mbox{  when  } p\rightarrow p+\lambda.$$
We denote the new bundle equipped with this affine structure, $T_B^*M$. Here the $b$ field is the component of the connection on an affine bundle corresponding to the translation subspace of the affine transformations.

Our new $b$ field immediately runs into a problem. The cohomology class of $db$ vanishes. To realize B fields with cohomologically nontrivial field strength, we pass to quotients of $T_B^*M$. There are two obvious ways to do this. The first
is to consider discrete quotients; then our fibers become products of tori and affine spaces. This cure allows the field strength of $b$ to lie in a discrete subgroup of $H^3(M)$ but restricts the geometry of $M$.

A second solution is given by quotienting by "gauge equivalence class". In other words, consider sections $p$ mod exact sections $df$. This does not give a finite dimensional bundle; so, to stay in the geometric regime, we consider the finite dimensional approximation to this equivalence given by 1 jets of sections quotiented by 1 jets of exact sections.  The resulting quotient space is an affine space locally modelled on $\bigwedge ^2T^*M$. If we allow affine changes of coordinates
$(x,p)\rightarrow (x,p+\lambda)$, where now $p$ and $\lambda$ are 2 forms, then we may define lifts of vector fields of the form
\begin{equation}\label{eq3}
\frac{\p}{\p x^i}\rightarrow \frac{\p}{\p x^i} + (b_{ji;k} + b_{ik;j} + c_{ijk} + p_{nk}\Gamma^n_{ij} + p_{jn}\Gamma^n_{ik})\frac{\p}{\p p_{jk}},
\end{equation}
where $b$ and $c$ are locally defined 2 and 3 forms respectively.
This is well defined if
$$b\rightarrow b-\lambda, \mbox{  and  }  c\rightarrow c+d\lambda, \mbox{  when  } p\rightarrow p+\lambda.$$
If we restrict to closed translations, $\lambda$, then we may choose $c=0$
and obtain b fields with arbitrary field strength without restricting the topology of $M$. For nonclosed $\lambda$ it is necessary to include the 3 form potential $c$. Once again, its field strength is cohomologically trivial in this formulation, and this may be remedied by passing to quotients. Discrete quotients constrain the geometry of $M$ and lead to field strengths of $c$ lying in discrete subgroups of $H^4(M)$. Gauge quotients lead to higher p-form potentials. We see that there is an analog of our p-form potential construction for all $p<dim M$.

We now see how these bundles are related to string theory. Let $s,t$ denote coordinates for a string with $t$ a timelike parameter and $s$ the position along the string. Let $X^{\mu}(s,t)$ denote coordinates for the string world sheet. We would like to study a quantum mechanical system which reflects some low energy information about the string sigma model. The simplest method to do this is to consider only the average value $x^{\mu}(t)$ of $X^{\mu}(s,t)$ and its conjugate momentum. This leads to the quantum mechanics of a point particle moving on the target space $M$.  This system loses too much string data. Motivated by the quantum mechanics of the (affine) 1 jet approximation of the string maps, we include the average value of $X_s^{\mu}$ in our system. The average value of $X_s^{\mu}$ is just $(X^{\mu}(\pi,t) - X^{\mu}(0,t))/\pi$. Jet space constructions suggest that we represent $X^{\mu}(\pi,t) - X^{\mu}(0,t)$ as a differential operator $\frac{1}{i}\frac{\partial}{\partial p_{\mu}}$ tangent to the fiber of an affine bundle, whose coordinates $p_{\mu}$ we  think of as velocities, or conjugate momenta to the $(X^{\mu}(\pi,t) - X^{\mu}(0,t))/\pi$.

Assuming the $p_{\mu}$ are coordinates for an affine fiber is clearly at best a low energy approximation. For example, if $X^{1}$ wraps a small circle then $X(s,t)$ will  not lie in a single coordinate chart for all $s$ and the average value of $X_s$ will also encode winding number (and only winding number for the closed strings). This dictates that in the $p_1$ direction, the noncompact affine fiber be replaced by a circle fiber dual to the wrapped circle, thus leading to discrete quotients of the affine fibers as required for cohomologically nontrivial B field field strengths. Then the commutator
$$[\frac{1}{i}\frac{\partial}{\partial p_{1}},e^{2\pi iwp_1}] = 2\pi we^{2\pi iwp_1}$$
with $\frac{1}{i}\frac{\partial}{\partial p_{1}}$ corresponding to $X^{1}(\pi,t) - X^{1}(0,t)$ suggests we interpret $p_1$ as the infinitesimal generator of motion along the circle. (Further deformations of the affine
geometry of the fibers which are suggested by supersymmetry are considered in \cite[Section 4.5]{S2}.)

 These considerations lead to the rough dictionary
$$\mbox{average  }X_s^{\mu}\rightarrow \frac{\partial}{\partial p_{\mu}},\mbox{          and}$$
$$\mbox{average  }X_t^{j}\mbox{ (or better -  } \pi^j) \rightarrow g^{ij}(\frac{\partial}{\partial x^i} + (\mu_{i;\nu} + b_{i\nu} + p_n\Gamma^n_{i\nu})\frac{\partial}{\partial p_{\mu}}.$$
Here $\pi^j$ denotes the total momentum in the $jth$ direction.
The interpretation of $p_1$ as a generator of motion in the $X^1$ direction further suggests we associate
$$X^{\mu}_s(\pi,t)\rightarrow p_{\mu}.$$
 
\section{D branes}
We now use the dictionary of the preceding section to see what form D branes must take in our model. An n-brane (in $M^n$) is given by fully Neumann boundary conditions: $X^{\mu}_s(0) = X^{\mu}_s(\pi) = 0$. From our dictionary, we see that this corresponds to a zero section (in a choice of local affine coordinates) of $T_B^*M$. A change in time parameter for the string world sheet $(s,t)\rightarrow (s,\tau(s,t))$ induces an affine change
$$\frac{\partial}{\partial s}\rightarrow \frac{\partial}{\partial s} + \frac{\partial\tau}{\partial s}\frac{\partial}{\partial \tau},$$
and correspondingly an affine change of coordinates in $T_B^*M$.  Thus the B field reflects the nonuniqueness of the time coordinate.

Consider next a p brane corresponding to a p dimensional submanifold $S$ of $M$. Choose local coordinates so that $S$ is given by  $x^{\nu}=0$ for $\nu > p$.  Then the $p-$brane boundary conditions $X^{\nu}_t = 0,$ $X^{\nu}_s$ free, for $\nu>p$ and $X^{\nu}_s = 0,$ $X^{\nu}_t$ free, for $\nu\leq p$ translate under our dictionary to
$x^{\nu}= 0$ and $p_{\nu}$ free for $\nu > p$ for some choice of local affine coordinates. This is an n dimensional submanifold $Z$ of $T_B^*M$ which, in a choice of local affine coordinates, is the conormal bundle of $S$. In particular, a zero brane is just an affine fiber over a point.

Fixing coordinates so that $Z$ is identified with a conormal bundle removes the gauge freedom to vary $\mu_{\nu}$ arbitrarily for $\nu \leq p$. Hence the D brane comes equipped with a locally defined 1 form $\mu$ on $S$. When the $B$ field is trivial, this defines the gauge field for a line bundle on $S$ (equipped with a local frame). On the overlap of two coordinate neighborhoods, $U_{\alpha}$ and $U_{\beta}$, we have $b^{\alpha} - d\mu^{\alpha} = b^{\beta} - d\mu^{\beta}.$ Thus, $\mu$ determines a connection on a line bundle only if the field strength of $b$ vanishes in $H^3(S).$  This is, of course, to be expected. In the presence of a B field with nontrivial field strength, the gauge field of a D brane is not that of a vector bundle, but of a "twisted bundle" (see for example \cite{W1},\cite{K},\cite{CKS}) or more generally perhaps of an infinite dimensional $C^*$ algebra \cite{BM}. Interpreting $\mu$ as a coordinate of the brane in a bundle is, perhaps, geometrically simpler than working with infinite dimensional $C^*$ algebra bundles.

If we have two distinct D branes $Z_1$ and $Z_2$ which correspond after distinct affine transformations to the normal bundle of a single submanifold $S$ of $M$, then we see that we have a gauge enhancement. In addition to the previously identified gauge fields, we also have the 1 form measuring the relative displacement in the fiber between the 2 branes.

In general, we would like to define a $D$ brane to be an n dimensional submanifold, $Z$ of $T_B^*M$ which corresponds to a choice of boundary condition, Dirichlet or Neumann, for each coordinate. If the Neumann boundary condition were well defined, then for $Z$ to represent a choice of boundary condition would imply that
the symplectic form $\omega:= dx^{\mu}\wedge dp_{\mu}$ vanishes when pulled back to $Z$. I.e., $Z$ is Lagrangian. Neither the form $\omega$ nor the Neumann condition is well defined. One can define a connection dependent analog of each, but  we will instead use the provisional definition:
\begin{define} A D brane is an n dimensional submanifold $Z$ of $T_B^*M$ such that for every point $p\in Z$ there exists a neighborhood of $p$ in $Z$ which, after an affine choice of coordinates can be identified with an open set in the conormal bundle of  some submanifold $S$ of $M$.
\end{define}

BPS conditions will restrict the possible $Z$ which occur.
This definition may be too broad. In particular, the distinction between $p$ brane and $p'$ brane for $p\not=p'$ becomes somewhat fuzzy. Nonetheless, it does provide a geometrical framework which includes sheaves. For example, assume now that $M$ is a complex manifold, and consider the ideal sheaf ${\mathcal I}_D$ of a divisor $D$. Let $z$ be a local defining function for $D$. The connection form $\frac{dz}{z}$ of ${\mathcal I}_D$ is singular along $D$. Hence the $n$ brane $Z_D$ given by the graph of $\frac{dz}{z}$ becomes vertical as it approaches $D$. This singularity cannot be removed by a (finite) affine change of coordinates and reflects the fact that ${\mathcal I}_D$ is not the sheaf of sections of a vector bundle. Assuming in this example that $T_B^*M = T^*M$, we may try to deform $Z_D$ to the zero section $Z_{\mathcal O}$, which is the $n$ brane corresponding to the trivial sheaf ${\mathcal O}$. Allowing only bounded affine shifts, we are left in the limit with
$$Z_D\rightarrow Z_{\mathcal O}\cup Z_{{\mathcal O}_D},$$
where $Z_{{\mathcal O}_D}$ is the brane corresponding to the sheaf
${\mathcal O}_D$. This gives a geometrical analog of the exact sequence of sheaves
$$0\rightarrow {\mathcal I}_D\rightarrow {\mathcal O}\rightarrow {\mathcal O}_D\rightarrow 0.$$

\section{Chern-Simons augmentation of the field strength}
In this section we show for heterotic strings how our treatment of B fields leads to the modification of the B field field strength by the addition of Chern-Simons terms.
We will only treat the Yang-Mills term in detail. Some of the material in this section postdate the conference and were elaborated at the suggestion of Eric Sharpe.

Let $\frakg$ be the Lie algebra of $SO(32)$ or $E_8\times E_8$.  Let $\frakh$ be a Cartan subalgebra. Let $P\subset \frakh^*$ denote the set of roots of $\frakg$ with respect to $\frakh$, and let $\{\alpha_i\}_{1\leq i\leq 16}$ be a basis of simple roots. Let $\Gamma$ denote the lattice in $\frakh^*$ generated by $P$. With respect to an inner product $(\cdot,\cdot)$ on $\frakh$ determined by the Killing form, $\Gamma$ is self dual and every element of $P$ has square length 2.

We use the bosonized description of the heterotic string. So, let $X$ be a 26 dimensional manifold which is a torus bundle with 10 dimensional base $M$ and fiber isometric to $\frakh/\Gamma$. Consider an affine cotangent bundle over $X$.  Quotient the affine fibers by the action generated by 1 forms dual to Killing vectors generating the torus lattice to obtain an affine bundle $T_{hetB}^*M$.

For $\alpha\in P$, let $\tau_{\alpha}$ denote the vector field on $\frakh/\Gamma$ which satisfies
$$\tau_{\alpha}\beta = (\alpha,\beta),\mbox{    for all  }\beta\in P.$$
Up to the familiar cocycle difficulty \cite[Section 6.4 Volume 1]{GSW} (most easily corrected by adjoining gamma matrix prefactors), the vector fields  $e^{2\pi i\alpha}\tau_{\alpha}, \alpha\in P \mbox{  and  }\tau_{\alpha_i}, 1\leq i\leq 16$ generate an algebra isomorphic to $\frakg$, when endowed with the Lie bracket given by their commutator composed with projection onto their span.

Let $B$ denote a $B-$field on $X$. We modify our prior assumption that $B$ is locally a 2 form pulled back from $X$ with the (chiral) assumption that it is allowed to vary in the $\frakh^*/\Gamma$ fiber but only (in an appropriate frame) as $B(x,t,s) = B(x,t+s)$,
where $x$ is a local coordinate on $M$, and $t$ and $s$ denote coordinates in $\frakh/\Gamma$ and $\frakh^*/\Gamma\simeq \frakh/\Gamma$ respectively. More precisely, we assume that $B$ has the form
$$B = b_{\nu\rho}dx^{\nu}\wedge dx^{\rho} + A_{\nu\alpha}e^{2\pi i\alpha(t+s)}dx^{\nu}\wedge d\alpha + A_{\nu i}dx^{\nu}\wedge d\alpha_i + \mbox{massive},$$
where other terms may occur but will be thought of as massive and ignored at this level.
Only those terms in $B$ which are invariant in the fiber and which are annihilated by interior multiplication by vectors tangent to the torus fiber descend to a 2 form on $M$. Thus, these lead to the "$B$ field" $b$ on $M$.

We also allow affine transformations $p\rightarrow p+\lambda$ to vary in the torus fiber;  we assume that $\lambda$ has the form
$$\lambda = \lambda^{-\alpha}(x)e^{2\pi i\alpha(t+s)}d\alpha + \lambda^i(x)d\alpha_i + \lambda^0(x) + \mbox{massive}.$$
Here $\lambda^0(x)$ is the pull back of a 1 form on $M$ and $\lambda^{\alpha}$ and $\lambda^i$ are the pull back of functions on $M$.
Under this transformation
$$b\rightarrow b + d\lambda^0 - A_{\mu\alpha}dx^{\mu}\wedge d\lambda^{-\alpha} - A_{\mu}^idx^{\mu}\wedge d\alpha_i(\lambda^{0})/2,$$
which under a natural identification can be written as
$$b\rightarrow b + d\lambda^0 + Tr A\wedge d\lambda.$$
Similarly we obtain a variation in $A$ which can be written as
$$A\rightarrow A+d(\lambda-\lambda^0) + [A,\lambda].$$
The new term in the variation of the $b$ (not B) field differs by an exact term from the usual Yang-Mills modification of the heterotic B field. This leads as usual to the Yang-Mills Chern-Simons modification of the field strength of $b$:
$$H = db -\omega_{YM},$$
but now $H$ arises simply as the component of $dB$ which descends to a form on $M$. Here $\omega_{YM}$ denotes the Yang-Mills Chern-Simons form.

To obtain the gravitational Chern-Simons form, $\omega_L$,  requires a bit more work, but seems likely to cast more light on the relation of heterotic string theory to M theory. The affine bundle must be replaced by a principal affine bundle. Then the modified field strength $$H = db + \omega_{L} - \omega_{YM}$$
again arises by taking the component of $dB$ which descends to $M$.

\section{Acknowledgements}
I would like to thank Ilarion Melnikov, Ronen Plesser, and Eric Sharpe for helpful discussions. This work was supported in part by NSF grant 0204188.


\begin{thebibliography}{99}


\bibitem{BM} P. Bouwknegt, V. Mathai, "D-branes, B fields, and twisted K-theory"
JHEP 0003 (2000) 007 hep-th/0002023.

\bibitem{CKS} A. Caldararu, S. Katz, and E. Sharpe, "D-branes, B fields and Ext groups" hep-th/0302099.

\bibitem{GSW} M. Green, J. Schwarz, and E. Witten, Superstring theory Volumes 1 and 2,  Cambridge, Cambridge University Press (1987).

\bibitem{H1} N. Hitchin, Lectures on special Lagrangian submanifolds,
{\em Winter School on Mirror Symmetry, Vector Bundles, and Lagrangian
Submanifolds}, 151-182, AMS/IP Stud Adv. Math. 23, Amer. Math. Soc., Providence, RI, 2001 DG/9907034.

\bibitem{H2} N. Hitchin, Generalized Calabi-Yau manifolds,  DG/0209099.

\bibitem{K} A. Kapustin, D-Branes in a topologically nontrivial B-field,
Adv. Theor. Math. Phys. 4 (2001) 127 hep-th/9909089.

\bibitem{OOY} H. Ooguri, Y. Oz, and Z. Yin, D-branes on Calabi-Yau Spaces and Their Mirrors, hep-th/9606112, Nuc. Phys. B 477 (1996) pp. 407-430.

\bibitem{Sh} E. Sharpe, D-Branes, Derived Categories, and Grothendieck Groups, hep-th/9902116, Nucl. Phys. B561 (1999) pp.433-450.

\bibitem{S1} M. Stern, Quantum mechanical mirror symmetry, D branes, and B
fields, hep-th/0209192.

\bibitem{S2} M. Stern, Mechanical D branes and B fields, hep-th/0310020.

\bibitem{W1} E. Witten, D branes and K theory, hep-th/9810188, JHEP 9812:025 (1998) 032.

\end{thebibliography}
\end{document}